\documentclass[twocolumn,showpacs,preprintnumbers,amsmath,amssymb]{revtex4}

\usepackage{graphicx}
\usepackage{dcolumn}
\usepackage{bm}
\usepackage{epstopdf}
\def\gsimeq{\,\,\raise0.14em\hbox{$>$}\kern-0.76em\lower0.28em\hbox {$\sim$}\,\,}
\def\lsimeq{\,\,\raise0.14em\hbox{$<$}\kern-0.76em\lower0.28em\hbox {$\sim$}\,\,}

\begin{document}

\preprint{PRC}

\title{Determination of the photodisintegration reaction rates involving charged particles: systematical calculations and proposed measurements based on Extreme Light Infrastructure - Nuclear Physics (ELI-NP)}

\author{H.Y. Lan$^{1}$, Y. Xu$^{2,\footnote{yi.xu@eli-np.ro}}$, W. Luo$^{1,\footnote{wenluo-ok@163.com}}$, D.L. Balabanski$^{2}$, S. Goriely$^{3}$, M. La Cognata$^{4}$, C. Matei$^{2}$, A. Anzalone$^{4}$, S. Chesnevskaya$^{2}$, G.L. Guardo$^{2}$, D. Lattuada$^{2}$, R.G. Pizzone$^{4}$, S. Romano$^{4,5}$, C. Spitaleri$^{4,5}$, A. Taffara$^{4}$, A. Tumino$^{4,6}$, Z.C. Zhu$^{1}$}

\affiliation{$^{1}$School of Nuclear Science and Technology, University of South China, 421001, Hengyang, China}
\affiliation{$^{2}$Extreme Light Infrastructure-Nuclear Physics, RO-077125, Magurele, Romania}
\affiliation{$^{3}$Institut d'Astronomie et d'Astrophysique, CP-226, Universite Libre de Bruxelles, 1050, Brussels, Belgium}
\affiliation{$^{4}$INFN-Laboratori Nazionali del Sud, 95123, Catania, Italy}
\affiliation{$^{5}$Department of Physics and Astronomy, University of Catania, 95123, Catania, Italy}
\affiliation{$^{6}$Kore University, 94100, Enna, Italy}

\date{\today}

\begin{abstract}
Photodisintegration reaction rates involving charged particles are of relevance to the p-process nucleosynthesis that aims at explaining the production of the stable neutron-deficient nuclides heavier than iron. In this study, considering the compound and pre-equilibrium reaction mechanisms, the cross sections and astrophysical rates of ($\gamma$,p) and ($\gamma$,$\alpha$) reactions for about 3000 target nuclei with $10 \le Z \le 100$ ranging from stable to proton dripline nuclei are computed. To study the sensitivity of the calculations to the optical model potentials (OMPs), both the phenomenological Woods-Saxon and the microscopic folding OMPs are taken into account. The systematic comparisons show that the reaction rates, especially for the ($\gamma$,$\alpha$) reaction, are dramatically influenced by the OMPs. Thus the better determination of the OMP is crucial to reduce the uncertainties of the photodisintegration reaction rates involving charged particles. Meanwhile, a $\gamma$-beam facility at Extreme Light Infrastructure-Nuclear Physics (ELI-NP) is being developed, which will open new opportunities to experimentally study the photodisintegration reactions of astrophysics interest. Considering both the important reactions identified by the nucleosynthesis studies and the purpose of complementing the experimental results for the reactions involving p-nuclei, the measurements of six ($\gamma$,p) and eight ($\gamma$,$\alpha$) reactions based on the $\gamma$-beam facility and the Extreme Light Infrastructure Silicon Strip Array (ELISSA) for the charged particles detection at ELI-NP are proposed. Furthermore, the GEANT4 simulations on these ($\gamma$,p) and ($\gamma$,$\alpha$) reactions are performed using the calculated cross sections and the features of the $\gamma$-beam facility and the ELISSA detector at ELI-NP. Simultaneously satisfying the minimum detectable limit of the experimental yield and the particle identification of proton and $\alpha$ particle, the minimum required energies of the $\gamma$-beam to measure the six ($\gamma$,p) and eight ($\gamma$,$\alpha$) reactions are estimated. It is shown that the direct measurements of these photonuclear reactions based on the $\gamma$-beam facility at ELI-NP within the Gamow windows at the typical temperature of T$_9$=2.5 for p-process are fairly feasible and promising. We believe that this pivotal work will guide the future photodisintegration experiments at ELI-NP. Furthermore, the expected experimental results will be used to constrain the OMPs of the charged particles, which can eventually reduce the uncertainties of the reaction rates for the p-process nucleosynthesis.
\end{abstract}


\maketitle

\section{Introduction}
\label{s1}

In massive star evolution and stellar explosive site, the astrophysical p-process \cite{AA1969,APJS1978,AA1990,AA1995,PR2003,APJ20031,APJ20032,APJ2008,APJ2014,IJMPE2016,MNRAS2016,MNRAS2018} is an important way of nucleosynthesis to produce the stable and proton-rich nuclei beyond iron which cannot be reached by the s- and r-processes. The common picture is that the p-nuclei are synthesized by photodisintegration of the pre-existing s- and r-processes nuclei via the ($\gamma$,n), ($\gamma$,p), and ($\gamma$,$\alpha$) reactions. The dominant nuclear flows of the p-process go towards the neutron-deficient region through the ($\gamma$,n) reactions. Along the isotopic paths, the proton separation energies become progressively smaller, while the corresponding energies for the neutrons go up. As a result, the flows to more neutron-deficient isotopes are hindered, and sometimes are deflected by the ($\gamma$,p) and ($\gamma$,$\alpha$) reactions. The typical temperature of the p-process is between T$_{9}$=1.5 (T$_{9}$ is the temperature in units of 10$^{9}$ K) and T$_{9}$=3.5.

For the complete determination of p-process, accurate knowledge of the capture and photodisintegration reaction rates for about 3000 stable and proton-rich nuclei is necessary. Although significant efforts have been made to study the reactions involved in the p-process, available experimental information, especially for the unstable and exotic nuclei, are still limited \cite{expsum2006,expsum2007,expsum2011,expsum2013,expsum2014}. To evaluate the reaction rates for which the experimental data are not available yet, state-of-the-art nuclear reaction models with the nuclear structure knowledge deduced by the microscopic models should be taken into account \cite{cal2004,cal2008,cal2012}. However, the theoretical estimates, for example within the
framework of the Hauser-Feshbach statistical model, remain uncertain due to the lack of complete nuclear structure knowledge, such like the nuclear potential of $\alpha$ particle at the energy region far below the Coulomb barrier \cite{apot1998,apot2013,apot2017}. On the other hand, research opportunities on the measurements of the photonuclear reactions \cite{npn2018} including the ($\gamma$,p) and ($\gamma$,$\alpha$) channels at the energy range of astrophysics interest, based on the new developed $\gamma$-beam facility at Extreme Light Infrastructure - Nuclear Physics (ELI-NP), are quite promising, which is expected to bring the new experimental constraint on the p-process nucleosynthesis \cite{elinpTDR}.

In the present study, the reaction mechanism and reaction model, as well as the OMPs for proton and $\alpha$ particle used for the calculations, are briefly described in Section \ref{s2}. The systematic computations of the cross sections and astrophysical rates for the ($\gamma$,p) and ($\gamma$,$\alpha$) reactions on about 3000 target nuclei are performed, and the results are presented in Section \ref{s3}, in which the measurements of the important ($\gamma$,p) and ($\gamma$,$\alpha$) reactions for the p-process nucleosynthesis are proposed as well. The realistic GEANT4 simulation for the measurements of these important reactions at the energy range of astrophysics interest based on the $\gamma$-beam facility at ELI-NP is conducted, and correspondingly the explicit results are given in Section \ref{s4}. Summary is presented in Section \ref{s5}.

\section{Theory}
\label{s2}
\subsection{Reaction mechanism and model}
\label{s21}

The compound nucleus contribution (CNC) is usually dominant to the reaction in the energy range of astrophysics interest, which is well described by the Hauser-Feshbach model \cite{HF1952}. This model relies on the fundamental Bohr hypothesis that the reaction occurs by means of the intermediary formation of a compound nucleus that can reach a state of thermodynamic equilibrium. The formation of a compound nucleus occurs if the nuclear level density (NLD) in the compound nucleus, at the excitation energy corresponding to the projectile incident energy, is sufficiently high \cite{nld1991,nld1996,nld1997}.

Provided the reaction $A+\gamma=B+b$ ($b$ = proton or $\alpha$ particle) represents the photon excitation on the nucleus $A$ resulting the residual nucleus $B$ with an emitted proton or $\alpha$ particle, the corresponding binary cross section of the CNC can be written as
\begin{eqnarray}
\sigma^{CNC}(E)=\sum^{A}_{x=0}\sum^{B}_{x=0}\sigma_{A^{x}+\gamma->B^{x}+b}^{CNC}(E).
\label{eqHF1}
\end{eqnarray}
\noindent
The summation $\sum^{A}_{x=0}$ and $\sum^{B}_{x=0}$, where the energy level scheme is represented by the $x$-th excited states ($x$=0 refers to the ground state), take over all the possible states (ground and excitation states) of the target $A$ and the residual nucleus $B$. Each state is characterized by its spin $S^{x}_{A}$, parity $\pi^{x}_{A}$ and excitation energy $E^{x}_{A}$ for the target $A$ (and similarly for the residual nucleus $B$).

The expression of the cross section $\sigma_{A^{x}+\gamma->B^{x}+b}^{CNC}(E)$ is given by (e.g.~\cite{cal2008,Xu2014})
\begin{eqnarray}
&&\sigma_{A^{x}+\gamma->B^{x}+b}^{CNC}(E_{\gamma})\nonumber\\
&&=\frac{\pi}{k^2}\sum^{l_{max}+S_{A^{x}}+S_{\gamma}}_{J=mod(S_{A^{x}}+S_{\gamma},1)}\sum^{1}_{\Pi=-1}
\frac{2J+1}{(2S_{A^{x}}+1)(2S_{\gamma}+1)}\nonumber\\
&&\times\sum^{J+S_{A^{x}}}_{\lambda=|J-S_{A^{x}}|}
\sum^{\lambda+S_{\gamma}}_{l_{i}=|\lambda-S_{\gamma}|}
\sum^{J+S_{B^{x}}}_{J_{b}=|J-S_{B^{x}}|}
\sum^{J_{b}+S_{b}}_{l_{f}=|J_{b}-S_{b}|}\nonumber\\
&&\times\delta^{\pi}_{C_\gamma}\delta^{\pi}_{C_b}\frac{\langle T^{J}_{C_\gamma,l_{i},\lambda}(E_\gamma)\rangle
\langle T^{J}_{C_b,l_{f},J_{b}}(E_b)\rangle}
{\sum_{Clj}\delta^{\pi}_{C}\langle T^{J}_{Clj}(E_{C})\rangle}
W^{J}_{C_{\gamma}l_{i}\lambda C_{b}l_{f}J_{b}},
\label{eqHF2}
\end{eqnarray}
\noindent
where $E_{\gamma}$ the incident energy of $\gamma$; $k$ the wave number of the relative motion; $l_{max}$ the maximum value of the relative orbital momentum; $J$ and $\Pi$ the total angular momentum and the parity of the compound nucleus; $S_{A^{x}}$ the spin of the target $A^{x}$, $S_{\gamma}$ the spin of photon, $\lambda$ the multi-polarity of photon (total angular momentum), $l_{i}$ the relative orbital momentum of the target $A^{x}$ and photon; $S_{B^{x}}$ the spin of the residual nucleus $B^{x}$, $S_{b}$ the spin of the emitted particle (proton or $\alpha$ particle here), $J_{b}$ the total angular momentum of the emitted particle, $l_{f}$ the relative orbital momentum of the residual nucleus $B^{x}$ and the emitted particle; $E_{b}$ the energy of the emitted particle; $C_{\gamma}$ the channel label of the initial system ($A^{x}$+$\gamma$) designated by $C_{\gamma}$=($\gamma$, $S_{\gamma}$, $E_{\gamma}$, $E_{A^{x}}$, $S_{A^{x}}$, $\pi_{A^{x}}$); $C_{b}$ the channel label of the final system ($B^{x}$+$b$) designated by $C_{b}$=($b$, $S_{b}$, $E_{b}$, $E_{B^{x}}$, $S_{B^{x}}$, $\pi_{B^{x}}$); $\delta^{\pi}_{C_{\gamma}}$=1 if $\pi_{A^{x}}$$\pi_{\gamma}$$(-1)^{l_{i}}$=$\Pi$ and 0 otherwise; $\delta^{\pi}_{C_{b}}$=1 if $\pi_{B^{x}}$$\pi_{b}$$(-1)^{l_{f}}$=$\Pi$ and 0 otherwise; $\pi_{\gamma}$ the parity of photon, $\pi_{b}$ the parity of the emitted particle; $T$ the transmission coefficient; $\sum_{Clj}\delta^{\pi}_{C}\langle T^{J}_{Clj}(E_{C})\rangle$ the sum of the transmission coefficient $T$ for all of the possible decay channels $C$ of the compound nucleus; and $W$ the width fluctuation correction factor for which different approximate expressions are described and discussed in Ref.~\cite{Hilaire2003}. In particular, the transmission coefficient for the particle emission is determined by the optical potentials between the two interacting particles, while the photon transmission coefficient relies on the $\gamma$-ray strength function.

The pre-equilibrium contribution (PEC) may become significant for the increasing energy or for the involved nuclei of which the compound nucleus does not have time to reach the thermodynamic equilibrium. The pre-equilibrium process can occur after the first stage of the reaction but long before the statistical equilibrium of the compound nucleus is reached. One of the most widely used model to describe the PEC is the (one- or two-component) exciton model \cite{Koning2004}, in which the nuclear state is characterized by, at any moment during the reaction, the total energy and the total number of particle($p$)-hole($h$) above and below the Fermi surface. Particles and holes are referred to as exciton. Furthermore, it is assumed that all possible ways of sharing the excitation energy between different particle-hole configurations at the same exciton number, $n=p+h$, have an equal a-priori probability. The basic starting point of the exciton model is a time-dependent master equation which describes the probability of the transitions to more and less complex $p$-$h$ states, as well as the transitions to the continuum (emission). The complete formalism of the exciton model can be found in Ref.~\cite{Koning2004}.

\begin{figure*}
\centering
\includegraphics[width=\textwidth]{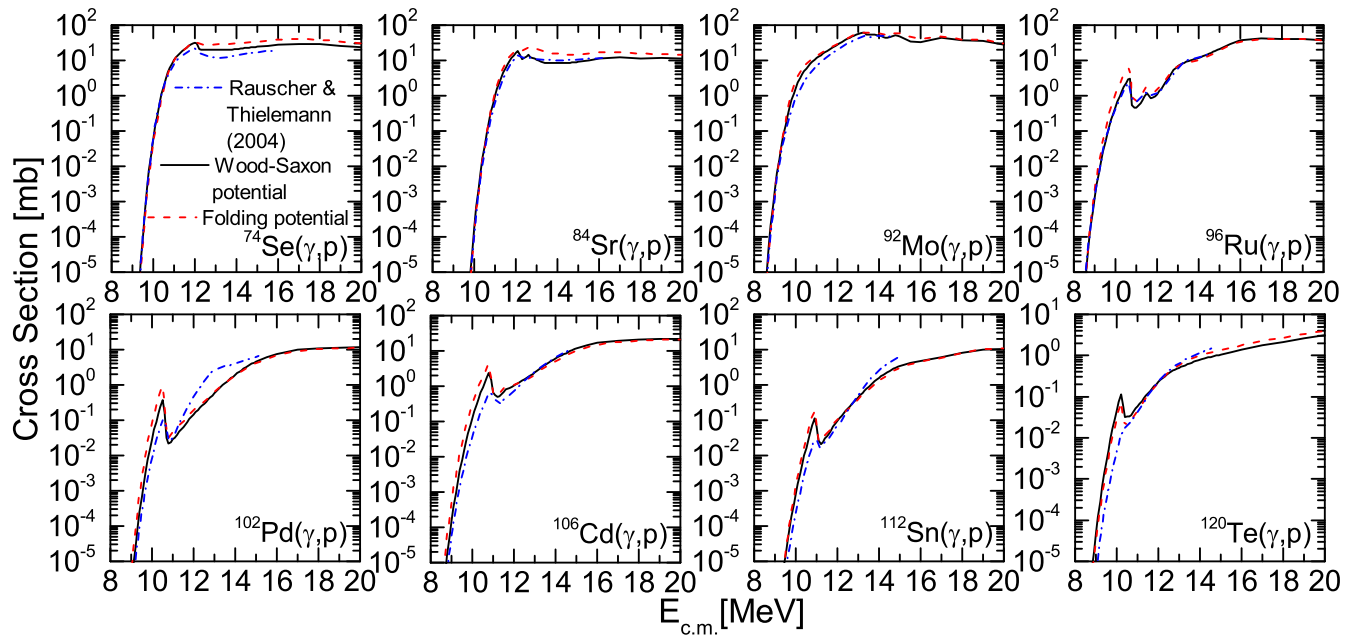}
\caption{Comparisons of the ($\gamma$,p) cross sections calculated by the phenomenological Woods-Saxon and the JLMB folding OMPs for eight p-nuclei targets of $^{74}$Se, $^{84}$Sr, $^{92}$Mo, $^{96}$Ru, $^{102}$Pd, $^{106}$Cd, $^{112}$Sn, and $^{120}$Te. The results calculated by the Woods-Saxon and the JLMB OMPs are respectively shown as the black solid lines and the red dash lines. The computations of Ref.~\cite{cal2004} are shown as the blue dash-dot lines for comparisons.}
\label{f1}       
\end{figure*}

\subsection{Optical model potentials (OMPs)}
\label{s22}

The nuclear ingredients, required for the reaction model calculation, can be extracted by the basic nuclear structure properties. Whenever available, the nuclear ingredients are taken from the experimental data, and if not, are deduced by the parametric phenomenological and microscopic models. The optical model potential (OMP) is an essential input for the calculations of the nuclear reaction properties, especially for the capture and photodisintegration reactions involving charged particles. In the present study, both of the parametric phenomenological and microscopic folding OMPs are employed to quantitatively and systematically study the cross sections and the reaction rates for ($\gamma$,p) and ($\gamma$,$\alpha$) reactions on about 3000 stable and proton-rich nuclei.

The Koning-Delaroche global phenomenological OMP with the Woods-Saxon form for the system of (nucleon + target) is described in detail in Ref.~\cite{WSp1}, which was completely verified by the extensive experimental data of nucleon induced reactions with the incident energies from 1 keV up to 200 MeV and the target masses range from $A$ = 24 to $A$ = 209. This OMP is based on a smooth and unique functional form for the energy dependence of the potential depths and the physically constrained geometry parameters.
The explicit expression reads
\begin{eqnarray}
U(E,r)&=&-V_{v}(E,r)-iW_{v}(E,r)-iW_{s}(E,r) \nonumber\\
&&+V_{s.o.}(E,r)+V_{c.}(r),
\label{ppot1}
\end{eqnarray}
\noindent
where $V_{v}$ and $W_{v,s}$ are the real and imaginary components of the volume-central (v) and surface-central (s) potentials; $V_{s.o.}$ is the spin-orbit potential; and $V_{c.}$ is the Coulomb potential, respectively. The central potentials are separated into energy-dependent well depths and energy-independent form factor, namely
\begin{eqnarray}
V,W_{v}(E,r)=V,W_{v}(E)\times f(r,R_{v},a_{v}),
\label{ppot2}
\end{eqnarray}
\noindent
and
\begin{eqnarray}
W_{s}(E,r)=-4a_{s}W_{s}(E)\times d(f(r,R_{s},a_{s}))/dr.
\label{ppot3}
\end{eqnarray}
\noindent
The form factor $f$ is in a Woods-Saxon form
\begin{eqnarray}
f(r,R_{i},a_{i})=(1+exp[(r-R_{i})/a_{i}])^{-1},
\label{ppot4}
\end{eqnarray}
\noindent
where the geometry parameters are the radius $R_{i}$ = $r_{i}A^{1/3}$ with $A$ being the atomic mass number and
the diffuseness $a_{i}$. The OMP parameterisation for proton in a function of $Z$, $A$ and incident energy $E$, deduced by the explicit expressions in Ref.~\cite{Koning2004}, is used for the present calculation.

\begin{figure*}
\centering
\includegraphics[width=\textwidth]{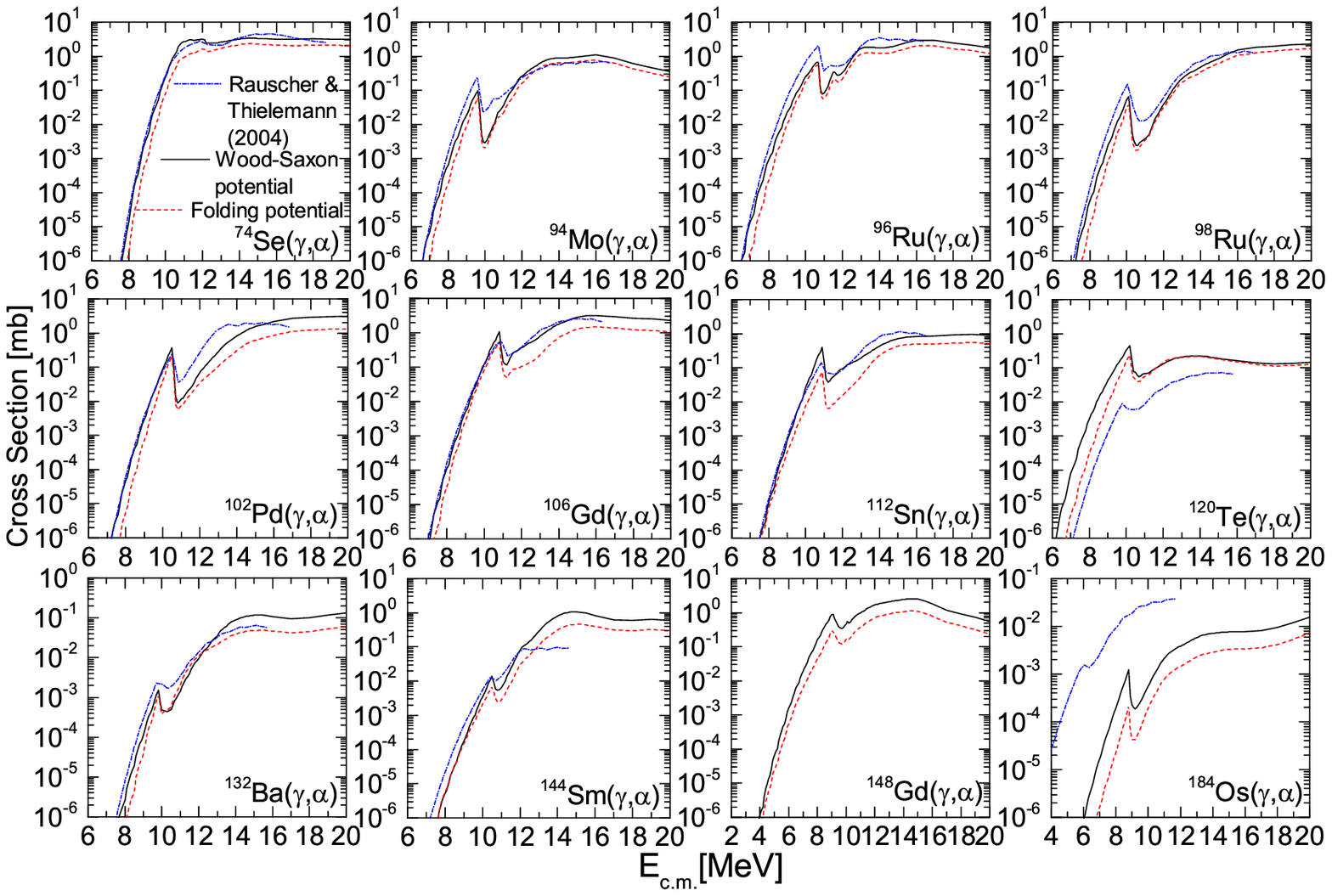}
\caption{Comparisons of the ($\gamma$,$\alpha$) cross sections calculated by the phenomenological Woods-Saxon and the M3Y folding potentials for twelve p-nuclei targets of $^{74}$Se, $^{94}$Mo, $^{96}$Ru, $^{98}$Ru, $^{102}$Pd, $^{106}$Cd, $^{112}$Sn, $^{120}$Te, $^{132}$Ba, $^{144}$Sm, $^{148}$Gd, and $^{184}$Os. The results calculated by the Woods-Saxon and the M3Y OMPs are respectively shown as the black solid lines and the red dash lines. The computations of Ref.~\cite{cal2004} are shown as the blue dash-dot lines for comparisons.}
\label{f2}       
\end{figure*}

On the other hand, the Bruy\`eres-le-Ch\^ atel renormalization \cite{JLMB2001} of the Jeukenne-Lejeune-Mahaux \cite{JLM1974,JLM1977}, referred as JLMB, is the global microscopic nucleon-nucleus OMP adjusted on the experimental data of the nuclei from $A$ = 30 to $A$ = 240  and for the energies ranging from 10 keV up to 200 MeV. The JLMB OMP has been phenomenologically renormalized in Refs.~\cite{JLMB2001,JLMB1998} to improve the agreement between the experimental observables and the predictions for a large set of data.

The JLMB OMP for a given nuclear matter density $\rho$ = $\rho_{n}+\rho_{p}$ and asymmetry $\alpha$ = ($\rho_{n}-\rho_{p}$)/$\rho$ reads
\begin{eqnarray}
V(E,r)&=&\lambda_{V}(E)[V_{0}(E)+\lambda_{V1}(E) \alpha V_{1}(E)]+ \nonumber\\
&&i\lambda_{W}(E)[W_{0}(E)+\lambda_{W1}(E) \alpha W_{1}(E)]
\label{ppot5}
\end{eqnarray}
\noindent
with $E$ the incident nucleon energy; $V_{0}(E)$, $V_{1}(E)$, $W_{0}(E)$, and $W_{1}(E)$ the real isoscalar, real isovector, imaginary isoscalar, and imaginary isovector components, respectively; $\lambda_{V}(E)$, $\lambda_{V1}(E)$, $\lambda_{W}(E)$, and $\lambda_{W1}(E)$ the respective renormalization factors. The explicit expressions can be found in Ref.~\cite{JLMB2001}.

The JLMB OMP has been extensively used for the nuclear astrophysics applications\cite{cal2008,Xu2014}, and here is also employed for the present calculations of the cross sections and the reaction rates. In particular, the matter density ($\rho_{n}$ and $\rho_{p}$) \cite{Goriely2010,BRUSLIB} predicted by the Hartree-Fock-Bogoliubov (HFB) method based on the BSk21 Skyrme interaction are used to calculate the four components of the JLMB potential in Eq.~(\ref{ppot5}) on the basis of local density approximation applied to the Br\"{u}ckner-Hartree-Fock calculation of the nuclear matter \cite{JLM1974,JLM1977}.

For the system of ($\alpha$ particle + target), the phenomenological Woods-Saxon OMP with the expressions of Eqs.~(\ref{ppot1}) and (\ref{ppot2}) are considered. The OMP parameters of depth, radius and diffuseness are determined by fitting the most available cross sections of $\alpha$ particle induced reactions below the Coulomb barrier on the target nuclei within the mass number of 45 $\leq$ $A$ $\leq$ 209 \cite{WSa1,WSa2}, and the explicit expressions of these parameters with $E$-, $A$- and $Z$- dependencies can be found in Refs.~\cite{WSa3,WSa4}. Since the experimental data below the Coulomb barrier are used to derive the potential parameters, this potential would be especially proper to calculate the cross sections and the reaction rates at the low energy range of astrophysics interest, and thus is used in the present calculation for the ($\gamma$,$\alpha$) reaction.

Meanwhile, global OMP for the system of ($\alpha$ particle + target) is proposed by Ref.~\cite{M3Y1}, taking into account the strong energy dependence and the nuclear structure effects characterizing the $\alpha$-nucleus interaction \cite{M3Y2}. The real part of this potential is obtained using a microscopic double-folding procedure over the M3Y effective nucleon-nucleon interaction, which can be described as:
\begin{eqnarray}
V_{DF}(E,r)=\int\int\rho_{p}(r_{p})\rho_{t}(r_{t}) \nonumber\\
\times\upsilon_{eff}(E,\rho=\rho_{p}+\rho_{t},s=|r+r_{p}-r_{t}|)d^{3}r_{p}d^{3}r_{t}.
\label{apot1}
\end{eqnarray}
\noindent
In Eq.~(\ref{apot1}), $\rho_{p}$ and $\rho_{t}$ are the density distributions of the projectile and the target, respectively; $r$ is the separation of the centers of mass of the target and the projectile; and $\upsilon_{eff}$ is the effective nucleon-nucleon interaction, which depends on the energy $E$ and the local densities $\rho_{p}$ and $\rho_{t}$. A phenomenological imaginary potential consisting of both the volume (Eq.~(\ref{ppot2})) and the surface (Eq.~(\ref{ppot3})) components with a Woods-Saxon form (Eq.~(\ref{ppot4})) is employed.

A significant improvement on this OMP of $\alpha$ particle is to import the dispersion relation linking the double-folding real parts with M3Y interaction and the phenomenological imaginary part with Woods-Saxon parameterization. Such additional constraint on the relation between the real part of the potential and the imaginary one can reduce the ambiguities in deriving the analytic expression of the potential from the experimental data. It is demonstrated that a large group of experimental data on the $\alpha$ particle elastic scattering and the $\alpha$ particle induced reactions at the energies of astrophysics application are well reproduced by the dispersive OMP. Global $\alpha$ particle OMP with the dispersion relation between the M3Y double-folding real part and the phenomenological Woods-Saxon imaginary part is used for the present calculation. The HFB-21 matter densities \cite{Goriely2010,BRUSLIB} are taken into account to generate the M3Y real part.

\section{Calculation results and suggestions for photonuclear measurements}
\label{s3}

\subsection{Systematical calculations}
\label{s31}

TALYS \cite{TALYS} is a software package for the simulation of nuclear reactions, which provides a complete description of all reaction channels and observables, and many state-of-the-art nuclear models covering all the main reaction mechanisms encountered in light particles induced nuclear reactions are included. The program is optimized for incident projectile energies ranging from 1 keV to 1 GeV on the target nuclei with mass numbers between 10 and 410. It includes photon, neutron, proton, deuteron, triton, $^{3}$He, and $\alpha$ particle as both projectiles and ejectiles, and single-particle as well as multi-particle emissions and fission. TALYS is designed to calculate the total and partial cross sections, the residual and isomer production cross sections, the discrete and continuum $\gamma$-ray production cross sections, the energy spectra, the angular distributions, the double-differential spectra, as well as the recoil cross sections.

For the ($\gamma$,p) and ($\gamma$,$\alpha$) reactions on about 3000 stable and proton-rich target nuclei, the calculations of the cross sections and the astrophysical reaction rates are performed with TALYS 1.8, which generate the results illustrated in Figures \ref{f1}-\ref{f4}. The nuclear structure ingredients used for the TALYS computations are explicitly presented as follows. The nuclear masses are taken from the Atomic Mass Evaluation 2016 (AME2016) \cite{AME} whenever available, while the HFB-21 nuclear masses \cite{Goriely2010} are taken into account when the AME2016 data are not available. The discrete experimental levels compiled in RIPL-3 library \cite{RIPL3} and the continuum level spectrum represented by the NLDs are both considered in the calculations of the photonuclear reactions. The NLDs are theoretically determined by the microscopic HFB plus a combinatorial approach \cite{NLD} that can well reproduce the low-lying cumulative number of the experimental levels. The photon strength functions obtained from the HFB plus quasiparticle random phase approximation (QRPA) \cite{SF} are used to calculate the electromagnetic transmission coefficients for the photon channel. The OMPs described in Section \ref{s22} are employed to determine the transmission coefficients for the particle channels. Specifically, the phenomenological Woods-Saxon OMPs \cite{WSp1} and the microscopic JLMB folding OMPs \cite{JLMB2001,JLM1974,JLM1977,JLMB1998} are respectively used for the calculations of the ($\gamma$,p) reactions, and two sets of results are correspondingly obtained. Similarly, for the ($\gamma$,$\alpha$) reactions, two sets of results are also computed using the phenomenological Woods-Saxon OMPs \cite{WSa3,WSa4} and the microscopic M3Y folding OMPs \cite{M3Y1}, respectively.

\subsection{Comparisons of the results}
\label{s32}

The comparisons of the ($\gamma$,p) cross sections calculated by the Woods-Saxon OMPs and the JLMB OMPs are shown in Figure \ref{f1} for eight p-nuclei of $^{74}$Se, $^{84}$Sr, $^{92}$Mo, $^{96}$Ru, $^{102}$Pd, $^{106}$Cd, $^{112}$Sn, and $^{120}$Te. It is found that at the energy range below $E_{c.m.}$ = 20 MeV, there is no dramatic disparity of the cross sections calculated by these two types of the OMPs. The comparisons of the ($\gamma$,$\alpha$) cross sections calculated by the Woods-Saxon OMPs and the M3Y OMPs are shown in Figure \ref{f2} for twelve p-nuclei of $^{74}$Se, $^{94}$Mo, $^{96}$Ru, $^{98}$Ru, $^{102}$Pd, $^{106}$Cd, $^{112}$Sn, $^{120}$Te, $^{132}$Ba, $^{144}$Sm, $^{148}$Gd, and $^{184}$Os. We can see that in Figure \ref{f2}, the differences of the ($\gamma$,$\alpha$) cross sections calculated by the two types of OMPs could reach one order of magnitude for most studied nuclei at the energy range of astrophysics interest. Since there is no direct measurement on the cross section of ($\gamma$,p) and ($\gamma$,$\alpha$) reactions below $E_{\gamma}$ = 20 MeV, we compare our theoretical results in Figures \ref{f1} and \ref{f2} to the available ($\gamma$,p) and ($\gamma$,$\alpha$) cross sections calculated by the Hauser-Fechbach model \cite{cal2004}. Fair agreements for most ($\gamma$,p) cross sections are found, while for ($\gamma$,$\alpha$) reactions, evident differences can be seen for the targets of $^{102}$Pd, $^{120}$Te and $^{184}$Os. The large discrepancy among the ($\gamma$,$\alpha$) cross sections shown in Figure \ref{f2} indicates that the OMPs for $\alpha$ particle, especially at the astrophysical energy range below Coulomb barrier, has not been determined very well, which could be the main source to cause the uncertainties of cross sections.

\begin{figure*}
\centering
\includegraphics[width=12cm,clip]{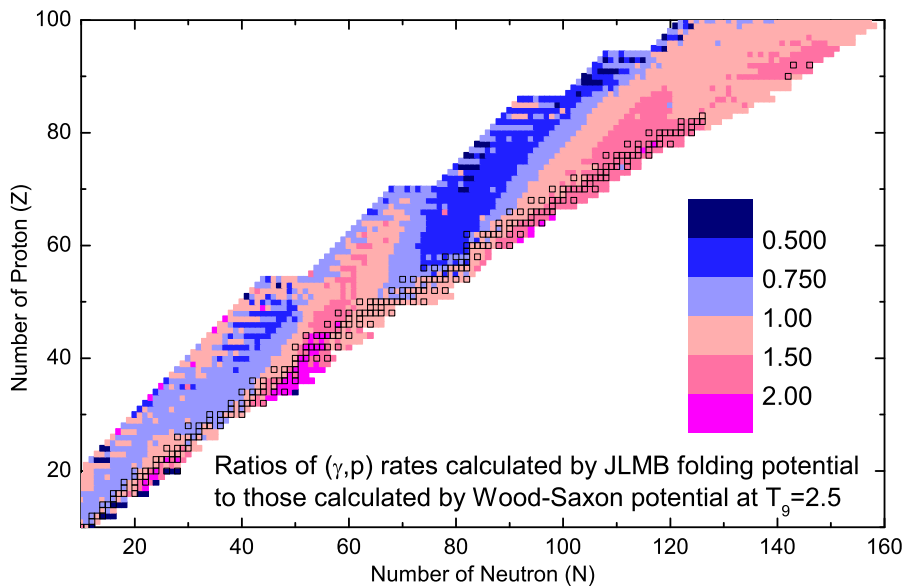}
\caption{Representation in the (N,Z) plane of the ratios of the ($\gamma$,p) astrophysical reaction rates at T$_{9}$=2.5 calculated by the microscopic folding JLMB OMPs to those calculated by the phenomenological Woods-Saxon OMPs for about 3000 targets of the stable and proton-rich nuclei.}
\label{f3}       
\end{figure*}

\begin{figure*}
\centering
\includegraphics[width=12cm,clip]{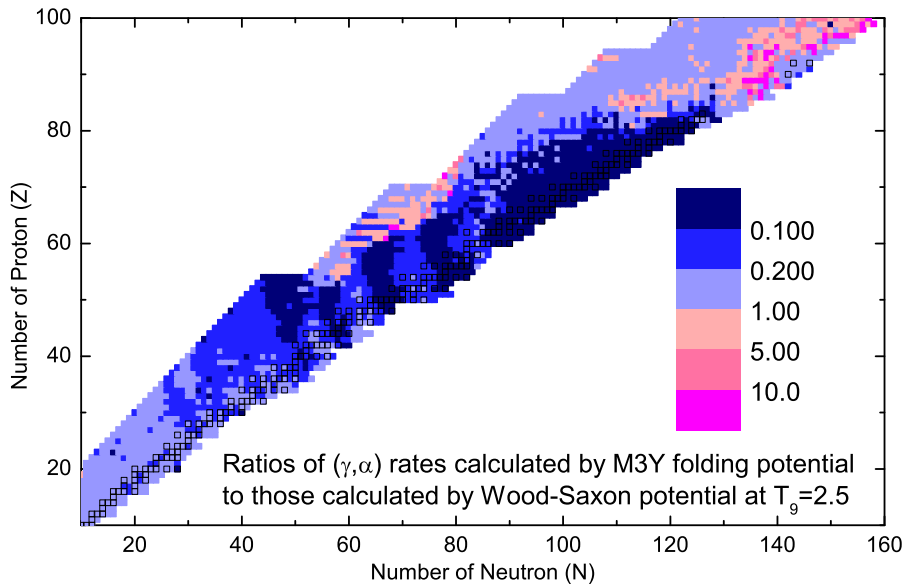}
\caption{Representation in the (N,Z) plane of the ratios of ($\gamma$,$\alpha$) astrophysical reaction rates at T$_{9}$=2.5 calculated by the microscopic folding M3Y OMPs to those calculated by the phenomenological Woods-Saxon OMPs for about 3000 targets of the stable and proton-rich nuclei.}
\label{f4}       
\end{figure*}

Furthermore, for the systematical investigation of the sensitivity of astrophysical reaction rate to the OMP, the comparisons of the ($\gamma$,p) reaction rates for about 3000 stable and proton-rich nuclei with $10 \le Z \le 100$, calculated by the microscopic folding JLMB OMP and the phenomenological Woods-Saxon OMP, are performed, and the similar comparisons are also carried out for the ($\gamma$,$\alpha$) reaction rates calculated by the microscopic folding M3Y OMP and the phenomenological Woods-Saxon OMP. Figure \ref{f3} and Figure \ref{f4} respectively represent in the ($N,Z$) plane the ratio of the ($\gamma$,p) and ($\gamma$,$\alpha$) reaction rates at T$_{9}$=2.5 calculated by the microscopic folding OMPs to those calculated by the phenomenological Woods-Saxon OMPs. It is demonstrated that the differences of ($\gamma$,$\alpha$) reaction rates could be one order of magnitude, especially for the proton-rich nuclei with $40 \le Z \le 80$ that can significantly contribute to the p-process nucleosynthesis.

\subsection{Proposed photonuclear experiments}
\label{s33}

The systematic comparisons reveal that the photodisintegration reaction rates involving charged particles, especially for the ($\gamma$,$\alpha$) reaction, can be dramatically influenced by the OMP. Therefore, the experimental studies of the photodisintegration reactions are proposed, and the OMP can be extracted and constrained by fitting the measured photonuclear data at the energy range of astrophysics interest, which are expected to be essentially used for the better determination of the reaction rates.

Although any new knowledge of the nuclear reactions involved in p-process is helpful, the experimental information of some specific photodisintegration reactions that can significantly affect the p-process nucleosynthesis is in particular desirable. Therefore, it is critical to identify the key photodisintegration reactions in the p-process nucleosynthesis, especially the ($\gamma$,p) and ($\gamma$,$\alpha$) reactions, that need prioritizing for the experimentally study. Recently, the uncertainties of the p-nuclei production originated by the variations of the astrophysical reaction rates are systematically studied, using the state-of-the-art Monte Carlo procedure \cite{MNRAS2016,MNRAS2018}, and the important reactions for determining the abundance of the p-nuclei are identified. In the present study, the key reactions of ($\gamma$,p) and ($\gamma$,$\alpha$) identified by Refs. \cite{MNRAS2016,MNRAS2018} are initially considered as the candidates of the photonuclear measurement. Furthermore, by explicitly checking the half-live of the nuclei involved in these candidate reactions, it is found that many of them cannot be experimentally studied via the photonuclear channel, due to the unavailability of the radiative targets. However, among these candidate reactions, the ($\gamma$,p) reaction on the target nuclei of $^{84}$Sr, $^{92}$Mo, $^{93}$Nb, and $^{96}$Ru, as well as the ($\gamma$,$\alpha$) reactions on the target nuclei of $^{74}$Se and $^{96}$Ru could be measured, if the availabilities of the targets are merely considered. These six reactions are selected for the simulation studies of the photonuclear measurements.

Besides the important reactions selected according to the astrophysical simulation, it is also worthwhile to conduct the measurements of some interesting ($\gamma$,p) and ($\gamma$,$\alpha$) reactions on the targets of p-nuclei. This is mainly because the inverse processes (e.g. the capture reactions) of such ($\gamma$,p) and ($\gamma$,$\alpha$) reactions have been experimentally studied, or cannot be measured due to the unavailability of the radiative targets. For example, the measurements of $^{70}$Ge($\alpha$,$\gamma$)$^{74}$Se \cite{capexp1}, $^{90}$Zr($\alpha$,$\gamma$)$^{94}$Mo \cite{capexp2}, $^{92}$Zr(p,$\gamma$)$^{93}$Nb \cite{capexp3}, $^{94}$Mo($\alpha$,$\gamma$)$^{98}$Ru \cite{capexp4}, and $^{108}$Cd($\alpha$,$\gamma$)$^{112}$Sn \cite{capexp5} have been performed, while the experiments of $^{73}$As(p,$\gamma$)$^{74}$Se, $^{111}$In(p,$\gamma$)$^{112}$Sn, $^{140}$Nd($\alpha$,$\gamma$)$^{144}$Sm, and $^{180}$W($\alpha$,$\gamma$)$^{184}$Os cannot be conducted because the radiative targets are not available. It is expected that the experiments of some interesting ($\gamma$,p) and ($\gamma$,$\alpha$) reactions on the targets of p-nuclei would be able to provide the significant supplements of the nuclear properties (e.g. the OMP) to the studies of p-process nucleosynthesis.

Therefore, considering both the important reactions identified by the nucleosynthesis studies and the purpose of complementing the experimental results for the interesting reactions involving p-nuclei, in the present study we eventually designate six ($\gamma$,p) reactions on the targets of $^{74}$Se, $^{84}$Sr, $^{92}$Mo, $^{93}$Nb, $^{96}$Ru, and $^{112}$Sn, and eight ($\gamma$,$\alpha$) reactions on the targets of $^{74}$Ge, $^{94}$Mo, $^{96}$Ru, $^{98}$Ru, $^{102}$Pd, $^{112}$Sn, $^{144}$Sm, and $^{184}$Os as the candidates for the further experimental campaigns. The realistic GEANT4 simulations of these fourteen photonuclear measurements at ELI-NP will be conducted in the following section.

\section{Simulation of photonuclear measurements based on ELI-NP}
\label{s4}

\subsection{$\gamma$-beam facility at ELI-NP}
\label{s41}

The Extreme Light Infrastructure-Nuclear Physics (ELI-NP) is aiming to use extreme electromagnetic fields for nuclear physics research \cite{elinp-ZamfirNuclPhysNews}, which comprises a high power laser system and a very brilliant $\gamma$-beam system. The technology involved in the construction of both systems is at the limits of the present-day technological capabilities. The high-brilliance narrow-bandwidth $\gamma$-beam, produced via Compton backscattering of a laser beam off a relativistic electron beam, will be delivered at ELI-NP, with the spectral density of $10^4$ photons/s/eV, the energies up to 19.5 MeV, and the bandwidth of 0.5$\%$. The main features of the $\gamma$-beam can be found in Ref.~\cite{elinp}. The research program of ELI-NP covers a broad range of the key topics in frontier fundamental physics and nuclear physics \cite{elinprev1,elinprev2}. In particular, thanks to the excellent features of the $\gamma$-beam, ELI-NP will provide unique opportunities to experimentally study the $\gamma$-induced reactions of nuclear astrophysics interest.

\subsection{Development of charged particles detector at ELI-NP}
\label{s42}

For the detection of charged particles, the silicon detector is one of the best solutions because it can guarantee an exceptional energy resolution and has essentially 100$\%$ efficiency. For the photonuclear reactions of astrophysical relevance, the energies of the emitted charged particles range from few hundreds keV to few MeV, so the low-threshold detector is necessary. Meanwhile, the kinematical identification is a viable option to separate the interesting particles from others. In practical, the silicon strip array has been successfully applied for the nuclear astrophysics studies, for example, the Oak Ridge Rutgers University Barrel Array (ORRUBA), the Array for Nuclear Astrophysics Studies with Exotic Nuclei (ANASEN) \cite{silicon}, the Advanced Implantation Detector Array \cite{AIDA}, and the silicon strip particle detector array TIARA \cite{TIARA}. With a common effort by ELI-NP and INFN-Laboratori Nazionali del Sud (INFN-LNS) in Catania, Italy, the Extreme Light Infrastructure Silicon Strip Array (ELISSA) is being developed \cite{elinpTDR}. The GEANT4 simulation \cite{dario2017} proves that a barrel configuration of ELISSA is particularly suited as it not only guarantees a very good resolution and granularity but also ensures a compact detection system and a limited number of electronics channel. The final design of ELISSA consists of three rings of twelve X3 position-sensitive detectors produced by Micron Semiconductor Ltd. \cite{Ltd} in a barrel-like configuration, with the assembly of four QQQ3 segmented detectors produced by Micron Semiconductor Ltd. as the end caps of both sides. Such configuration ensures a total angular coverage of $20\leq\theta\leq160$ in the laboratory system. Furthermore, the prototype of ELISSA has been constructed and tested at INFN-LNS with the $\alpha$ particle source and the 11 MeV $^{7}$Li beam, and the preliminary experimental results show that the energy resolution is less than 1$\%$ and the position resolution reaches 1 mm \cite{marc2017,luca2017,svet2017,svet2018}.

\subsection{GEANT4 simulation: Algorithm}
\label{s43}

The experiments of the six ($\gamma$,p) reactions and eight ($\gamma$,$\alpha$) reactions identified in Section \ref{s33} are simulated with the help of GEANT4-GENBOD approach \cite{luo2017}, which is a data-based Monte Carlo program aim to accurately simulate the specific photonuclear reactions in the framework of GEANT4. The ($\gamma$,p) and ($\gamma$,$\alpha$) cross sections calculated by the microscopic folding OMPs in Section \ref{s31} are incorporated in the simulation as the inputs, to generate the spectra and experimental yields of the emitted charged particles. In the simulation, the features of the $\gamma$-beam facility at ELI-NP with the photon intensity of $10^4$ photons/s/eV and the energy bandwidth of 0.5$\%$ \cite{elinp} are taken into account, and the configuration of ELISSA \cite{elinpTDR} is implemented accordingly. A double-layer structure of the target is used for the simulation, consisting of a 10 $\mu$m target foil facing the $\gamma$-beam and a 0.266 $\mu$m Carbon backing.

Besides the photonuclear reactions, the incident $\gamma$-beam can further induce Compton effect and pair production in the target, which is also taken into account by invoking the electromagnetic physical process in the simulation. In order to separate the interesting charged particles from others, discrimination should be made on the energy spectra of the outgoing particles including electron, photon, proton and $\alpha$ particle. In fact, the background event rate of Compton effect and pair production is rather small. The energy deposit of the electron background is as low as several hundreds keV, which can be readily removed by introducing a negligible threshold on the detector. Therefore, the key issue is to distinguish the products from the photonuclear nuclear reactions. The possibility to use the kinematical method for disentangling the charged particles emitted from the photodisintegration reactions of astrophysical relevance will be investigated in the following parts.

\subsection{GEANT4 simulation: Spectra}
\label{s44}

\begin{figure*}[!htb]
\centering
\includegraphics[width=15cm,clip]{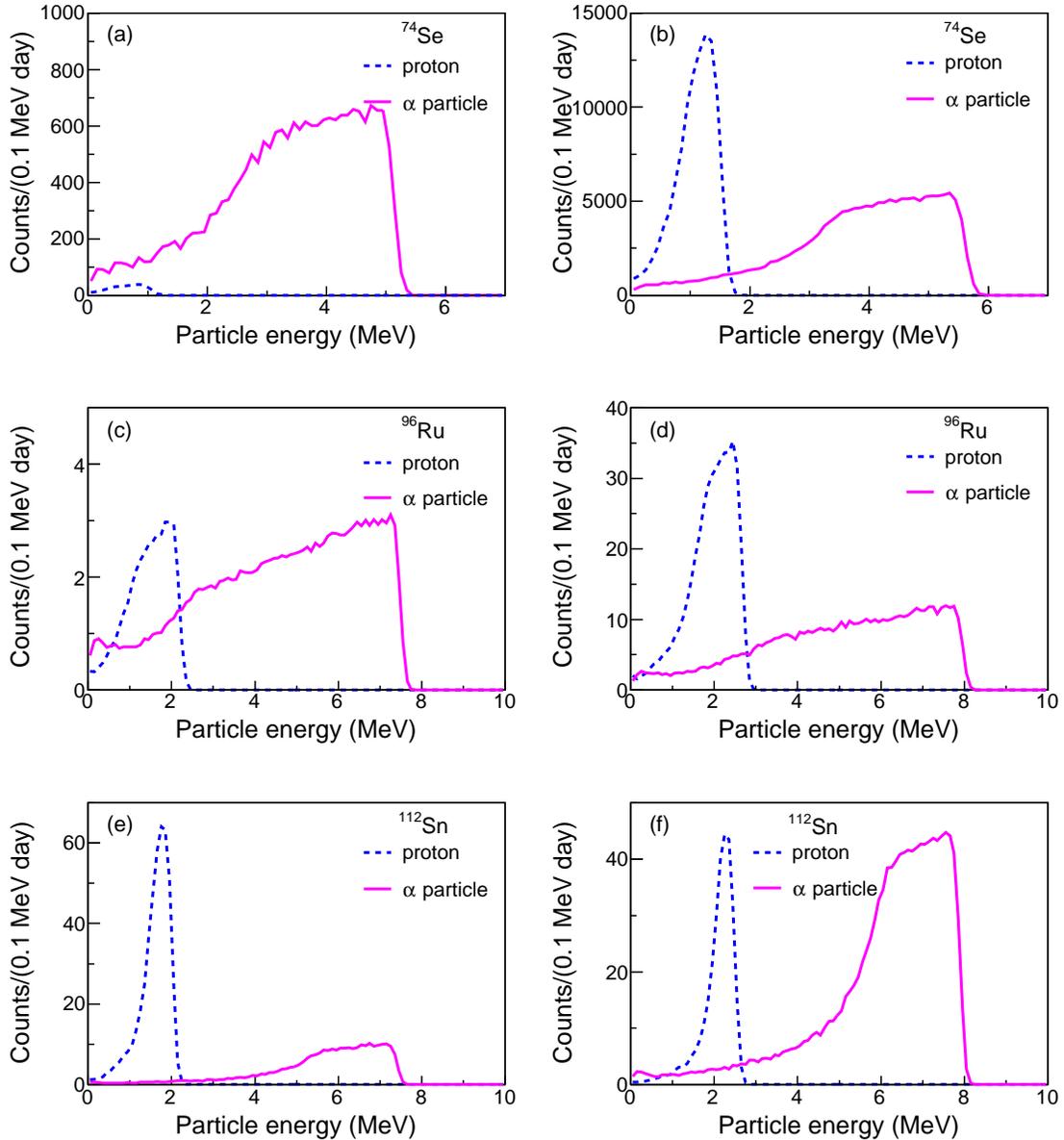}
\caption{The energy spectra of $\alpha$ particle (solid line) and proton (dash line) from the photodisintegration reactions on the targets of $^{74}$Se, $^{96}$Ru, and $^{112}$Sn at E$_{\gamma}$ = 9.5 MeV [(a), (c) and (e)] and E$_{\gamma}$ = 10 MeV [(b), (d) and (f)], respectively.}
\label{f5}       
\end{figure*}

For a given photonuclear reaction A($\gamma$,b)B, the peak energy of the emitted particle can be calculated by the kinematical equation of
\begin{eqnarray}
E_b = \frac{M_B}{M_B+M_b}(E_{\gamma}+Q),
\label{simu}
\end{eqnarray}
where \emph{M$_{B}$} and \emph{M$_{b}$} are the atomic mass of the residual nucleus and the emitted particle, respectively. In the present case \emph{b} denotes proton or $\alpha$ particle, \emph{E$_{\gamma}$} is the energy of the incident $\gamma$-beam, \emph{Q} is the Q-value of the reaction A($\gamma$,b)B. It can be seen that from Eq.~(\ref{simu}), the peak energy of the emitted proton is generally different to that of the emitted $\alpha$ particle, due to the discrepancy between the Q-value of the ($\gamma$,p) reaction and that of the ($\gamma$,$\alpha$) reaction. Such difference, as an energy gap, can be expressed by
\begin{eqnarray}
\Delta E &=& E_{\alpha}- E_{p}  \nonumber\\
&=& \frac{M_{R}}{M_{R}+M_{\alpha}}(E_{\gamma}+Q_{(\gamma,\alpha)}) \nonumber\\
&&- \frac{M_{R}}{M_{R}+M_{p}}(E_{\gamma}+Q_{(\gamma,p)}).
\label{simu1}
\end{eqnarray}
The existence of \emph{$\Delta$E} is beneficial to the disentanglement of the charged particles, though the ELISSA detector already ensures a high energy resolution as mentioned above. Obviously, the larger energy gap (\emph{$\Delta$E}) indicates the better particle identification between proton and $\alpha$ particle.

It is worth noting that in principle the particle disentanglement can be performed based on the kinematics of the emitted particles. However, the theoretical calculations show that the angular dependence of the ($\gamma$,p) and ($\gamma$,$\alpha$) reactions is very weak, and the nearly isotropic angular distributions of proton and $\alpha$ particle are obtained due to the dominant contribution from the compound reaction mechanism at the low energy range of astrophysics interest. Therefore, the entire energy spectra remains as an effective way of particle identification.

\begin{figure}[!htb]
\includegraphics[width=8.2cm,clip]{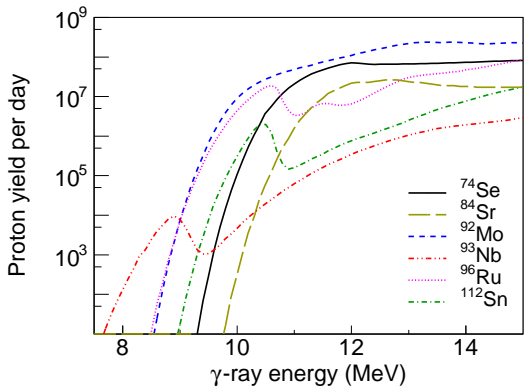}
\caption{The proton yield (count per day) of the ($\gamma$,p) reaction on the targets of $^{74}$Se, $^{84}$Sr, $^{92}$Mo, $^{93}$Nb, $^{96}$Ru, and $^{112}$Sn. The thickness of target is set as 10 $\mu$m.}
\label{f6}       
\end{figure}

\begin{figure*}[!htb]
\centering
\includegraphics[width=16.5cm, clip]{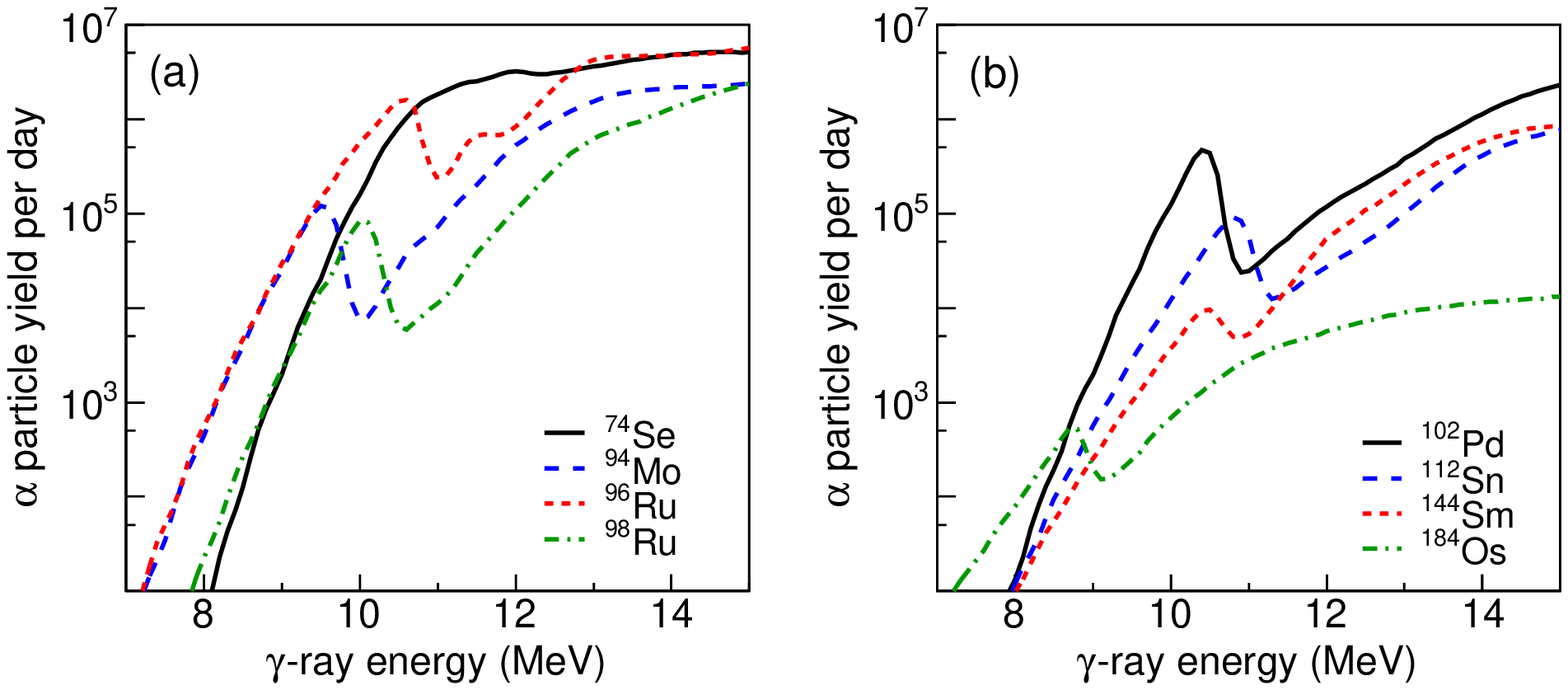}
\caption{The $\alpha$ particle yield (count per day) of the ($\gamma$,$\alpha$) reaction on the targets of $^{74}$Se, $^{94}$Mo, $^{96}$Ru, $^{98}$Ru, $^{102}$Pd, $^{112}$Sn, $^{144}$Sm, and $^{184}$Os. The thickness of target is set as 10 $\mu$m.}
\label{f7}       
\end{figure*}

The simulated energy spectra of proton and $\alpha$ particle emitted from the $\gamma$-induced reactions on $^{74}$Se, $^{96}$Ru and $^{112}$Sn are obtained and shown in Figure~\ref{f5}. Two different $\gamma$-induced energies, 9.5 MeV and 10 MeV, are used for the simulation. The energy gap \emph{$\Delta$E} for the ($\gamma$,p) and ($\gamma$,$\alpha$) reactions on $^{74}$Se is approximately 4.5 MeV, and for those reactions on $^{96}$Ru and $^{112}$Sn, \emph{$\Delta$E} increases to $\sim$ 5.5 MeV. The energy gaps obtained by the simulation are in good agreement with the values estimated via Eq.~(\ref{simu1}), which reveals the reasonability of the present simulation.

Meanwhile, Figure~\ref{f5} shows that the proton energies are much lower than the $\alpha$ particle energies, due to the larger separation energies of proton. Furthermore, the energy spectra of proton are visibly narrower than those of $\alpha$ particle, because of the longer energy range and the weaker energy loss of proton in the target. In Figure~\ref{f5} we can see that at \emph{E$_{\gamma}$} $\leq$ 10 MeV, the peak energy of proton is as low as 1.0 $-$ 3.5 MeV, and the peak energy of $\alpha$ particle is in the range of 5.0 $-$ 8.0 MeV. This means that $\alpha$ particle can be readily identified in the measurement. The background of $\alpha$ particle generated by the photonuclear reaction on the Carbon backing can hardly impact the measurement, because the cross section of $^{nat.}$C($\gamma$,$\alpha$) is much less than those of the studied targets. On the other hand, compared to the case of $\alpha$ particle, the proton is potentially likely to be contaminated by the background events of electron generated by the $\gamma$-beam interaction with the target foil and Carbon backing. However, this kind of background event rate is rather small \cite{luo2017}.

\subsection{GEANT4 simulation: Reaction yields}
\label{s45}

The total reaction yield at a given incident $\gamma$-beam energy can be obtained by integrating the energy spectra. At the incident $\gamma$-beam energy range of 8 $-$ 15 MeV, the simulated ($\gamma$,p) reaction yields for the six targets of $^{74}$Se, $^{84}$Sr, $^{92}$Mo, $^{93}$Nb, $^{96}$Ru, and $^{112}$Sn are shown in Figure~\ref{f6}, and the simulated ($\gamma$,$\alpha$) reaction yields for the eight targets of $^{74}$Se, $^{94}$Mo, $^{96, 98}$Ru, $^{102}$Pd, $^{112}$Sn, $^{144}$Sm, and $^{184}$Os are shown in Figure~\ref{f7}, respectively. In general, the experimental yields of both proton and $\alpha$ particle drop dramatically with the decrease of the incident $\gamma$-beam energy. For the targets of $^{74}$Se, $^{96}$Ru, and $^{112}$Sn, the $\alpha$ particle yields are lower by one order of magnitude than the proton yields at the same incident $\gamma$-beam energy.

The required energies of the $\gamma$-beam which can satisfy the minimum measurable limit of the experimental yield is necessary to evaluate the feasibility of the photonuclear measurement. In the present simulation, the minimum measurable limit for proton or $\alpha$ particle is set as 100 counts per day. We choose such relative larger amount of the minimum measurable limit (100 counts per day) in order to account for the uncertainties of the theoretical cross sections, as the experimental yields are simulated based on the theoretical cross sections. To meet this criteria of the detectable limit, the minimum required energies of the $\gamma$-beam for the ($\gamma$,p) and ($\gamma$,$\alpha$) reactions are deduced. For the six ($\gamma$,p) reactions of $^{74}$Se, $^{84}$Sr, $^{92}$Mo, $^{93}$Nb, $^{96}$Ru, and $^{112}$Sn, the minimum required energies of the $\gamma$-beam (E$_{\gamma}^{th}$) are 9.6, 10.3, 9.3, 8.1, 9.2, and 9.4 MeV, respectively. That is to say that, the ($\gamma$,p) cross sections down to several nanobarns can be measured. For the ($\gamma$,$\alpha$) reactions of $^{74}$Se, $^{94}$Mo, $^{96}$Ru, and $^{98}$Ru, the minimum required energies of the $\gamma$-beam (E$_{\gamma}^{th}$) are 8.5 MeV, 7.7 MeV, 7.8 MeV, and 8.5 MeV, respectively. The ($\gamma$,$\alpha$) reactions of $^{102}$Pd, $^{112}$Sn, and $^{144}$Sm share approximately the same E$_{\gamma}^{th}$ = 8.6 MeV, and E$_{\gamma}^{th}$ for the $^{184}$Os($\gamma$,$\alpha$) reaction is as low as 8.3 MeV.

\subsection{Discussion of the photonuclear measurements}
\label{s46}

A more important issue is to estimate the minimum required energies of the inducing $\gamma$-beam for the measurements of the ($\gamma$,p) and ($\gamma$,$\alpha$) reactions, simultaneously satisfying both the minimum measurable limit of the experimental yield (100 count per day) and the particle disentanglement between proton and $\alpha$ particle. At first, we note that the threshold energy of the ($\gamma$,$\alpha$) reaction is usually lower than that of the ($\gamma$,p) reaction. Therefore, below the threshold energy of the ($\gamma$,p) reaction, only $\alpha$ particle can be found if the experimental yield reaches the detectable criteria of 100 count per day. In this situation, no particle identification needs to be performed, which is the advantage for the $\alpha$ particle detection.

When the $\gamma$-beam energy rises above the threshold energy of the ($\gamma$,p) reaction, proton would be detected if the experimental yield reaches the detectable limit. However, such experimental event of proton could not be identified due to the overlap spectra of proton and $\alpha$ particle. For example, in Figure~\ref{f5}(a), the spectra of proton can not be separated from the spectra of $\alpha$ particle at E$_{\gamma}$ = 9.5 MeV for $^{74}$Se($\gamma$,p) reaction, hence the experimental yield can no longer be counted accurately. It is known from Section \ref{s44} that, the minimum required energy of the $\gamma$-beam for $^{74}$Se($\gamma$,p) reaction that meets the measurable limit of 100 count per day is 9.6 MeV. However, according to the simulation result for the energy spectra, such minimum energy has to be moved upward by 0.3 MeV if the particle disentanglement is taken into account simultaneously. Similarly, in order to satisfy the measurable criteria including both the minimum detectable limit of experimental yield (100 count per day) and the particle identification between proton and $\alpha$ particle, the minimum required energy of the $\gamma$-beam for $^{84}$Sr($\gamma$,p) reaction raises by 0.3 MeV approximately, and for $^{96}$Ru($\gamma$,p) and $^{112}$Sn($\gamma$,p) reactions it raises by 0.2 MeV. For $^{92}$Mo($\gamma$,p) and $^{93}$Nb($\gamma$,p) reactions, such minimum required energy of the $\gamma$-beam remains the same value given by Section \ref{s44}. Note that because of the extremely low cross section for $^{92}$Mo($\gamma$,$\alpha$) reaction, the experimental yield of $\alpha$ particle would not influence the detection of proton around and above the incident $\gamma$-beam energy of 9.3 MeV.

\begin{figure*}[!htb]
\centering
\includegraphics[width=18cm,clip]{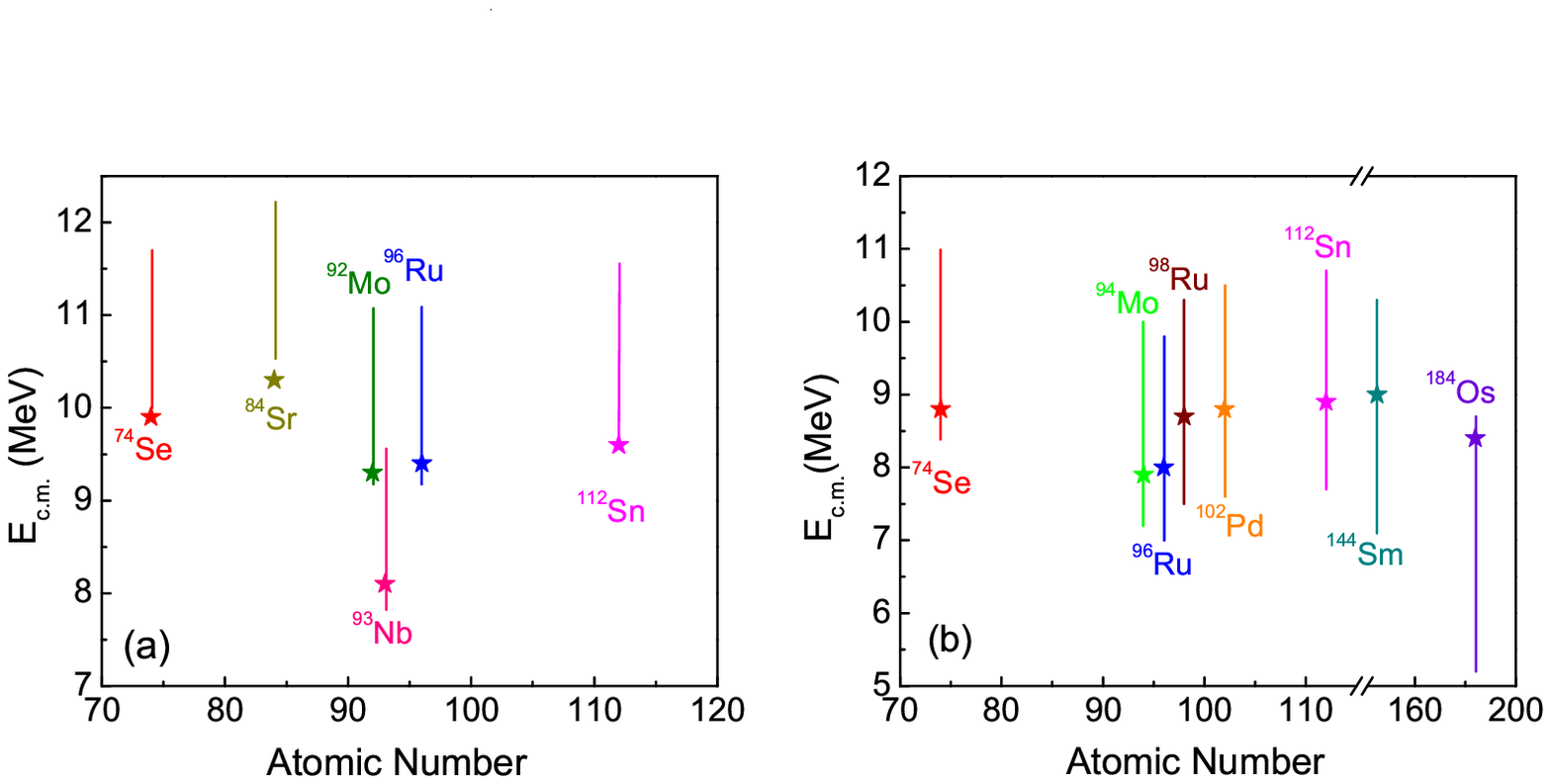}
\caption{The Gamow windows (shown as the length of the lines) at $T_{9}$=2.5 and the minimum required energies of the incident $\gamma$-beam (solid pentagrams) satisfying the measurable criteria of both the minimum detectable limit (100 count per day) and the particle identification for (a) the ($\gamma$,p) reactions on the targets of $^{74}$Se, $^{84}$Sr, $^{92}$Mo, $^{93}$Nb, $^{96}$Ru, and $^{98}$Ru, and (b) the ($\gamma$,$\alpha$) reactions on $^{74}$Se, $^{94}$Mo, $^{96}$Ru, $^{98}$Ru, $^{102}$Pd, $^{112}$Sn, $^{144}$Sm, and $^{184}$Os.}
\label{f8}       
\end{figure*}

Beyond the minimum required energy of the $\gamma$-beam for the ($\gamma$,p) reactions described as above, the experimental yields of proton and $\alpha$ particle are comparable. In this energy range, thanks to the sufficiently large gap (\emph{$\Delta$E} introduced in Section \ref{s43}) between the energy spectra of proton and $\alpha$ particle, it is possible to perform the particle identification.
In general, both the ($\gamma$,p) and ($\gamma$,$\alpha$) reactions can be measured simultaneously, when the energy of the inducing $\gamma$-beam is larger than the proton separation energy by about 1.0 - 2.0 MeV. This means that the minimum detectable cross section is in the order of 10$^{-4}$ mb.

The feasibility to perform the experiments of the six ($\gamma$,p) reactions on $^{74}$Se, $^{84}$Sr, $^{92}$Mo, $^{93}$Nb, $^{96}$Ru, and $^{98}$Ru as well as the eight ($\gamma$,$\alpha$) reactions on $^{74}$Se, $^{94}$Mo, $^{96}$Ru, $^{98}$Ru, $^{102}$Pd, $^{112}$Sn, $^{144}$Sm, and $^{184}$Os directly in their respective Gamow windows \cite{expsum2007,gamow,gamowgamma} are further evaluated. For these reactions, the Gamow windows at $T_9$ = 2.5, as a typical temperature of the p-process, are calculated and illustrated by the length of the lines in Figure~\ref{f8}. Meanwhile, the minimum required energies of the inducing $\gamma$-beam that satisfy the measurable criteria including both the minimum detectable limit (100 counts per day) of the reaction yields and the particle identification are correspondingly shown in Figure~\ref{f8} by the pentagram points. For the ($\gamma$,p) reactions in Figure~\ref{f8}(a), the Gamow windows locate above the minimum required energies of the inducing $\gamma$-beam. This means that the measurements of these ($\gamma$,p) reactions can be readily conducted in the energy ranges of the Gamow windows at $T_9$ = 2.5. In Figure~\ref{f8}(b) it is shown that the $^{74}$Se($\gamma$,$\alpha$) reaction can be measured in the entire energy range of the Gamow window at $T_9$ = 2.5. However, the experiments of the rest ($\gamma$,$\alpha$) reactions could be conducted only in 10$\%$ - 70$\%$ of the entire Gamow windows at $T_9$ = 2.5 (the upper parts of the lines above the pentagram points shown in Figure~\ref{f8}), due to the insufficient experimental yields of these ($\gamma$,$\alpha$) reactions below the minimum required energies of the inducing $\gamma$-beam. According to this feasibility investigation, it is suggested that the experiments of $^{74}$Se($\gamma$,$\alpha$) and $^{92}$Mo($\gamma$,p) reactions based on the $\gamma$-beam facility at ELI-NP could be given precedence.

So far, only the ground state of the residual nucleus generated by the photodisintegration is taken into account in the present simulation and discussion. This is because the contribution from the photodisintegration occurring to the ground state of the residual nucleus is generally dominant, when the inducing $\gamma$-beam energy is relatively lower (e.g., a few MeV higher than the particle separation energy). However, the proportion of the produced residual nuclei in their excited states may be considerable, when the inducing energy of the $\gamma$-beam increases. In this case, it is important to further investigate the possibility to disentangle the reaction products in different final states, which can be practically done, for example, by distinguishing the emitted $\alpha$ particle from the ($\gamma$,$\alpha$$_{0}$) and ($\gamma$,$\alpha$$_{1}$) channels. Such relevant studies of the photonuclear reactions involving the excited states are in progress, which would allow us to better discriminate the influence of the OMP, apart from other nuclear properties, on the reaction yields.

\section{Summary}
\label{s5}

The photodisintegration reaction rates involving charged particles are of relevance to the p-process nucleosynthesis that aims at explaining the production of the stable and neutron-deficient nuclides heavier than iron observed up to now in the solar system exclusively. In the present study, considering the compound and pre-equilibrium reaction mechanisms, we compute the cross sections and the astrophysical reaction rates for the ($\gamma$,p) and ($\gamma$,$\alpha$) reactions on about 3000 target nuclei with $10 \le Z \le 100$ ranging from the valley of $\beta$-stability to the proton drip line. Furthermore, the phenomenological Woods-Saxon and microscopic folding OMPs are both used in the present calculation to investigate the sensitivity of the reaction rate to the OMP. According to the systematic comparisons of the present calculations and the previous results, fair agreements of the studied ($\gamma$,p) cross sections are obtained, but the evident differences of the cross sections for the ($\gamma$,$\alpha$) reactions on the targets of $^{102}$Pd, $^{120}$Te, and $^{184}$Os are found. Furthermore, for about 3000 target nuclei, the ratios of the reaction rates computed by the Woods-Saxon OMP to those computed by the folding OMP at T$_{9}$=2.5 are represented on the N-Z plane, which reflects that the ($\gamma$,$\alpha$) reaction rates, especially for the proton-rich nuclei with $40 \le Z \le 80$ that significantly contribute to the p-process, are very sensitive to the OMP of $\alpha$ particle. Therefore, it is revealed that the better determination of OMP, especially at the energy range below the Coulomb barrier, is crucial to reduce the uncertainties of the photodisintegration reaction rates involving charged particles.

A new $\gamma$-beam facility at ELI-NP is being developed, and it will open new opportunities for experimentally studying the photodisintegration reactions of astrophysics interest. Furthermore, the development of the charged particles detector ELISSA allows us to measure the photodisintegration reactions involving charged particles. Considering both the important reactions identified by the nucleosynthesis studies and the purpose of complementing the experimental results for the reactions involving p-nuclei, the measurements of six ($\gamma$,p) reactions on $^{74}$Se, $^{84}$Sr, $^{92}$Mo, $^{93}$Nb, $^{96}$Ru, and $^{98}$Ru, as well as eight ($\gamma$,$\alpha$) reactions on $^{74}$Se, $^{94}$Mo, $^{96}$Ru, $^{98}$Ru, $^{102}$Pd, $^{112}$Sn, $^{144}$Sm, and $^{184}$Os are proposed based on the $\gamma$-beam facility and the ELISSA detector at ELI-NP. In particular, the GEANT4 simulations for these ($\gamma$,p) and ($\gamma$,$\alpha$) reactions of astrophysics interest are conducted using the calculated cross sections and the features of ELI-NP and ELISSA, and the energy spectra and reaction yields of the emitted charged particles are subsequently obtained. Moreover, taking into account the measurable criteria of both the minimum detectable limit for the experimental yield (100 counts per day) and the particle identification for proton and $\alpha$ particle, the minimum required energies of the inducing $\gamma$-beam to measure the six ($\gamma$,p) and the eight ($\gamma$,$\alpha$) reactions are estimated, which locate within their respective Gamow windows at T$_9$=2.5. Therefore, it is quite feasible and prospective to conduct the experiments of these proposed ($\gamma$,p) and ($\gamma$,$\alpha$) reactions directly within the energy ranges of the Gamow windows at T$_9$=2.5. Eventually, we expect that the present pivotal work will guide the future measurements of the photodisintegration reactions at ELI-NP. The future experimental results will be used to constrain the OMPs of the charged particles, which can reduce the uncertainties of the reaction rates for the p-process nucleosynthesis.

\begin{acknowledgements}
The authors gratefully thank the anonymous reviewers for the insight comments and suggestions. This work is supported by the Extreme Light Infrastructure Nuclear Physics (ELI-NP) Phase II, a project co-financed by the Romanian Government and the European Union through the European Regional Development Fund - the Competitiveness Operational Programme (1/07.07.2016, COP, ID 1334), the National Natural Science Foundation of China Grant No. 11675075, and the Natural Science Foundation of Hunan Province, China (Grant No. 2018JJ2315). W. L. appreciates the support from Youth Talent Project of Hunan Province, China (Grant No. 2018RS3096).
\end{acknowledgements}

\end{document}